\documentclass[a4paper,twocolumn]{esapub} 
\usepackage[sectionbib]{natbib}
\usepackage{graphicx}
\usepackage{xspace}
\usepackage{times}
\usepackage[dvips]{color}

\newcommand{\e}[1]{\ifmmode{%
\mathchoice{^{\mbox{\scriptsize #1}}}{^{\mbox{\scriptsize #1}}}%
{^{\mbox{\tiny #1}}}{^{\mbox{\tiny #1}}}}%
\else{$^{\mbox{\scriptsize #1}}$}\fi}
\renewcommand{\i}[1]{\ifmmode{%
\mathchoice{_{\mbox{\scriptsize #1}}}{_{\mbox{\scriptsize #1}}}%
{_{\mbox{\tiny #1}}}{_{\mbox{\tiny #1}}}}%
\else{$_{\mbox{\scriptsize #1}}$}\fi}
\newcommand{\ei}[2]{\ifmmode{%
\mathchoice{^{\mbox{\scriptsize #1}}_{\mbox{\scriptsize #2}}}%
{^{\mbox{\scriptsize #1}}_{\mbox{\scriptsize #2}}}%
{^{\mbox{\tiny #1}}_{\mbox{\tiny #2}}}%
{^{\mbox{\tiny #1}}_{\mbox{\tiny #2}}}}%
\else{$^{\mbox{\scriptsize #1}}_{\mbox{\scriptsize #2}}$}\fi}

\newenvironment{shortlist}{\begin{list}{--}%
{\leftmargin 2em\labelwidth 1em\labelsep 1em \parsep 0.2\baselineskip
\itemsep 0pt\topsep 0.2\baselineskip}}{\end{list}}

\newcommand{\ev}{e\kern -0.11em V\xspace}
\newcommand{\kev}{ke\kern -0.09em V\xspace}
\newcommand{\keV}{\kev}

\newcommand{\ca}{\mbox{$\sim$}}
\newcommand{\ariel}{\textsl{Ariel~5}\xspace}
\newcommand{\asca}{\textsl{ASCA}\xspace}
\newcommand{\heao}{\textsl{HEAO~1}\xspace}
\newcommand{\einstein}{\textsl{Einstein (HEAO~2)}\xspace}
\newcommand{\macho}{\textsl{MACHO}\xspace}
\newcommand{\rosat}{\textsl{ROSAT}\xspace}
\newcommand{\sax}{\textsl{BeppoSAX}\xspace}
\newcommand{\xmm}{\textsl{XMM-Newton}\xspace}
\newcommand{\afive}{\mbox{A\,0538--66}\xspace}
\newcommand{\sxj}{\mbox{SAX\,J1808.4--3658}\xspace}
\newcommand{\exo}{\mbox{EXO\,1745--248}\xspace}

\newcommand{\redchi}{\ensuremath{\chi\ei{2}{red}}\xspace}

\newcommand{\nh}{\ensuremath{N_{\rm H}}\xspace}

\newlength{\tabwidth}
\begin{document}

\title{XMM-Newton Observations of the Be/X-ray transient A0538-66 in quiescence}

\author[1,5]{P. Kretschmar}
\author[2,3]{J. Wilms}
\author[3]{R.~Staubert}
\author[3,5]{I. Kreykenbohm}
\author[4]{W.A.~Heindl}
\affil[1]{Max-Planck-Institut f\"ur Extraterrestrische Physik, 
                 87548 Garching, Germany}
\affil[2]{Department of Physics, University of Warwick, 
                 CV4~7AL Warwick, UK}
\affil[3]{Institut f\"ur Astronomie und Astrophysik -- Astronomie,
                 Univ. of T\"ubingen, 72076 T\"ubingen, Germany}
\affil[4]{Center for Astrophysics and Space Science, UCSD, 
                 La Jolla, CA, USA}
\affil[5]{INTEGRAL Science Data Center, 1290 Versoix, Switzerland}

\maketitle


\begin{abstract}
   We present XMM-Newton observations of the recurrent Be/X-ray
   transient A0538-66, situated in the Large Magellanic Cloud, in the
   quiescent state. Despite a very low luminosity state of 
   (5-8)$\times$10\e{33} ergs/s in the range 0.3-10\,\kev, 
   the source is clearly detected up to \ca 8\, keV. and can be fitted
   using either a power law with photon index $\alpha$=1.9$\pm$0.3 or
   a bremsstrahlung spectrum with $kT$=3.9\ei{+3.9}{$-$1.7}\,\kev.
   The spectral analysis confirms that the off-state spectrum is
   hard without requiring any soft component, contrary to the majority
   of neutron stars observed in quiescence up to now.
\end{abstract}

\section{Introduction} 

The recurrent Be/X-ray transient \afive (X\,0535$-$668) was discovered in
1977 when two outbursts were observed with the \ariel satellite
\citep{WhiteCarpenter:78}. In the following years several other outbursts
were observed with \heao and \einstein 
\citep[e.g.][]{Johnston:80,Skinner:80_Xray} with coincident X-ray and
optical flares showing a recurrence of 16.65 days \citep{Skinner:80_optical}
interpreted as the period of an eccentric binary orbit.
This period has been confirmed as a by-product of the \macho project
by \citet{Alcock:2001}, who also found a longer optical modulation of 421~days.
The optical counterpart, identified by \citet{Johnston:80}, was
found to be a member of the Large Magellanic Cloud (LMC) 
\citep{PakullParmar:81}. Optical and UV spectroscopy \citep{Charles:83} 
classify the counterpart as B2~IIIe star.

Based on the distance to the LMC, the luminosity of the outbursts observed
in the first years after detection can be estimated as around 10\e{39} ergs/s
making this a super-Eddington source and one of the most powerful X-ray
binaries known. An observation with \einstein 1980/81 during a strong
outburst found 69\,ms pulsations \citep{Skinner:82,Ponman:84}; up to
this day this remains the only measure of the pulse period of this
accreting pulsar. From the rapid spin period and the luminosity 
\citet{Skinner:82} inferred an upper limit to the magnetic field of 
$B$\ca 10\e{11}\,G assuming that matter accreted onto the polar caps
uninhibited by a centrifugal barrier.

In later years the source has been mostly quiescent.
Weak outburst activity at two to three orders of magnitude below
the early observations was found in the \rosat All-Sky Survey
in 1990 \citep{MavromatakisHaberl:93} and in an \asca observation
in 1995 \citep{Corbet:97}; for details see Table~\ref{Tab:Fits}.
\citet{Campana:2002} observed the source in 1999 in quiescence with \sax
finding very different results than in the previous observations.

\afive was observed in the quiescent state in April 2002 by \xmm.
The goal of the observation was to use the unique collecting power of
this satellite in order to obtain spectral data allowing to distinguish 
between different models of the quiescent emission, e.g., residual
accretion to the surface or accretion stopped at the magnetosphere.

\begin{figure*}[ht]

\centerline{\includegraphics[angle=-90,width=0.98\textwidth]{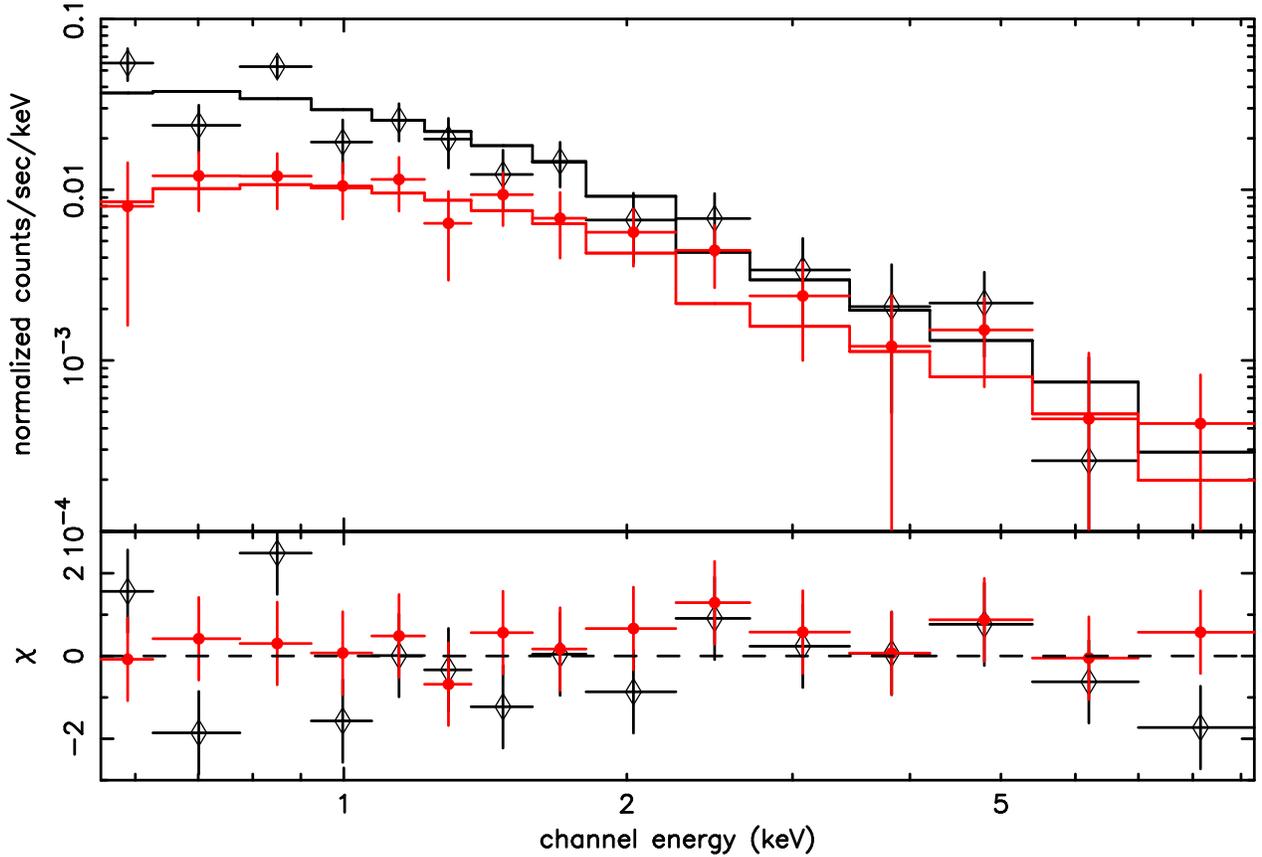}}

\caption{\label{Fig:Spec}Observed \xmm spectra for single (black, $\diamond$) 
         and double
         (\textcolor[named]{Red}{red, $\bullet$}) events, together with the best fit 
         power law spectrum.}

\end{figure*}

\section{Data Analysis}
\label{data}

The data were extracted with the \xmm \textsl{Scientific Analysis
Software}, version~5.4 from circles centered on the source, and from a
background region with the same area. Separate spectra were extracted for
single and double events, which allows to gauge potential calibration
uncertainties.

of the used spectral models. The amount of
photoelectric absorption is not strongly constrained and consistent
with zero. We used a fixed value of
\nh=8$\times$10\e{20} cm$^{-2}$, as found by \citet{Corbet:97} and
consistent with the ranges given in \citet{MavromatakisHaberl:93} and
\citet{Campana:2002}. 
\pagebreak[4]

The spectra can be well fitted either by a power law with photon index
$\alpha$=1.9$\pm$0.3 and normalization 
$I\i{1 keV}$=(4.9$\pm$0.6)$\times$10\e{$-$5}
ph\,keV\e{$-$1}\,cm\e{$-$2}\,s$^{-1}$ (\redchi$=$0.98) 
or a bremsstrahlung spectrum with temperature
$kT$=3.9\ei{+3.9}{$-$1.7}\,\kev and normalization
$I$=6.3\ei{+1.1}{$-$0.8}$\times$10\e{$-$5}
ph\,keV\e{$-$1}\,cm\e{$-$2}\,s$^{-1}$ (\redchi$=$1.03).  
A blackbody spectrum is not consistent with the data, 
showing clear systematic deviations above \ca 2\, keV (\redchi$=$1.86) .

The spectrum reported by \citet{Corbet:97} for a weak outburst (see
also Tab.~\ref{Tab:Fits}, first spectrum) is not at all compatible
with our data -- even after allowing for free scaling,
\redchi$\approx$6.75 for 29 d.o.f.  We can exclude any significant
contribution from an additional soft component or an iron line to the
best fit spectra -- at most such components could contribute 2--3\% to
the total flux.  In summary our results match very well those obtained
by \citet{Campana:2002}, but with the source in an even lower
luminosity state.

A period search by epoch folding in the period range
10--100\,ms found no pulsations in our dataset in line
with previous observations at higher flux levels.

\begin{table*}

\caption{\label{Tab:Fits}Comparison of spectral fit results with previous 
         observations of \afive. While in the outbursts soft spectra or
         spectral components were observed, the available data in quiescence
         shows hard spectra without the need for any soft component.
         Over the years the maximum observed luminosities have decreased.
         The luminosities here are given for a distance of 52\,kpc.}

\renewcommand{\arraystretch}{2.0}
\begin{center}
  \begin{tabular}{@{}lcrc}
   Authors & Time of Observation     & Luminosity [ergs/s] & Energy range \\
  \hline
  \rule{0pt}{2.5\parskip}\citet{Ponman:84} & Dec 1980 \& Feb 1981 & (2-5)$\times$10\e{38} & 1.5--20\,\kev \\
  \multicolumn{4}{p{\tabwidth}}{
        blackbody with $kT$=2.4\,\kev \textsl{or} \newline  
        Wien law with  $kT$=2.4\,\kev
  } \\
  \hline
  \rule{0pt}{2.5\parskip}\citet{MavromatakisHaberl:93} & Nov 1990 & 4$\times$10\e{37} & 
        0.1--2.4\,\kev \\[3pt]
  \multicolumn{4}{p{\tabwidth}}{
        blackbody with $kT$=0.25$\pm$0.04\,\kev \textsl{or} \newline
        bremsstrahlung with $kT$=0.9\ei{+0.9}{$-$0.3}\,\kev \textsl{or} \newline
        power law with $\alpha$=2.6$\pm$0.7
  } \\
  \hline
  \rule{0pt}{2.5\parskip}\citet{MavromatakisHaberl:93} & Dec 1990 & 2$\times$10\e{37} & 
        0.1--2.4\,\kev \\[3pt]
  \multicolumn{4}{p{\tabwidth}}{
        blackbody with $kT$=0.22$\pm$0.02\,\kev \textsl{or} \newline
        bremsstrahlung with $kT$=0.7\ei{+0.1}{$-$0.2}\,\kev \textsl{or} \newline
        power law with $\alpha$=2.9$\pm$0.3
  } \\
  \hline
  \rule{0pt}{2.5\parskip}\citet{Campana:97} & Oct 1993 & 1$\times$10\e{36} & 
        0.1--2.4\,\kev \\[3pt]
  \multicolumn{4}{p{\tabwidth}}{
        blackbody with $kT$=0.22\ei{+0.07}{$-$0.04}\,\kev \textsl{or} \newline
        power law with $\alpha$=2.5+1.2-0.9
  } \\
  \hline
  \rule{0pt}{2.5\parskip}\citet{Corbet:97} & Feb 1995 &  5$\times$10\e{36} & 0.3--10\,\kev \\[3pt]
  \multicolumn{4}{p{\tabwidth}}{
	power law ($\alpha$=2.04$\pm$0.15) plus 
        blackbody ($kT$=2.69$\pm$0.15\,\kev) 
        and Fe line at  6.52$\pm$0.04\,\keV, \textsl{or}\newline
	power law ($\alpha$=2.05$\pm$0.15) plus 
        blackbody ($kT$=2.69$\pm$0.15\,\kev) 
        and Fe lines at 6.4 and 6.7\,\keV, \textsl{or}\newline
        power law ($\alpha$=0.73$\pm$0.05) plus bremsstrahlung 
        ($kT$=0.58$\pm$0.35\,\kev) 
        and Fe line at 6.52$\pm$0.04\,\keV 
  } \\
  \hline
  \rule{0pt}{2.5\parskip}\citet{Campana:2002} & Sep 1999 & (1-3)$\times$10\e{35} & 0.1--10\kev \\[3pt]
  \multicolumn{4}{p{\tabwidth}}{
	power law with $\alpha$=2.1$\pm$0.6, \textsl{or}\newline  
        bremsstrahlung with $kT$=5.5\ei{+9.6}{$-$2.9}\,\kev 
  } \\
  \hline\hline
  \textcolor[named]{Red}{\rule{0pt}{2.5\parskip}\textbf{This study}} & 
  \textcolor[named]{Red}{\textbf{April 2002}} & 
  \textcolor[named]{Red}{\textbf{(5-8)$\times$10\e{33}}} & 
  \textcolor[named]{Red}{\textbf{0.3--10\,\kev}} \\[3pt]
  \multicolumn{4}{p{\tabwidth}}{\textcolor[named]{Red}{
	power law with $\alpha$=1.9$\pm$0.3, \textsl{or}\newline
	bremsstrahlung with $kT$=3.9\ei{+3.9}{$-$1.7}\,\kev} 
  } \\
  \hline
  \end{tabular}
\end{center}

\end{table*}

\section{Discussion}

The observations reported here found \afive in a state of very low
luminosity -- with (5-8)$\times$10\e{33} ergs/s this is the lowest
state for which spectral results have been reported up to today. 
The results found by \citet{Campana:2002} for the quiescent state --
i.e., a relatively hard spectrum which can be described as a power law
of photon index \ca 2 and no sign of pulsations
-- are confirmed and extended to even lower luminosity.

Several mechanisms to produce
the residual luminosity of an accreting X-ray pulsar in quiescence
are proposed by different authors:

\begin{shortlist}
\item Residual accretion onto the neutron star surface at very low
      rates \citep[e.g.][]{Stella:94}.
\item Accretion down to the centrifugal barrier caused by the rotating
      magnetic field \citep{IllarionovSunyaev:75} with the main emission 
      coming from
      the material outside or just at the magnetosphere boundary
      \citep{CampanaStella:2000}.
\item Thermal emission of the neutron star surface due to cooling of
      the core heated during phases of strong accretion 
      \citep{BrownBildstenRutledge:98}.
\item Rotational energy release via an activated radio pulsar
      either with a directly observed pulsar wind or X-ray emission
      originating in a shock front between the pulsar wind and the
      inflowing matter \citep[e.g.][]{Burderi:2002,Campana:2002}.
\end{shortlist}

From an observational point of view two groups of sources emerge:
The majority of the reported results has a significant soft component,
usually modeled as a blackbody spectrum, and sometimes a high energy
tail, represented by a power law of photon index $\alpha$ \ca 1--2 
\citep[e.g.][]{Asai:98}. Only three neutron star systems have been observed 
with a quiescent spectrum dominated by hard power law: 
\sxj \citep{Campana:2002}, \exo \citep{Wijnands:astro-ph}
and \afive. 
It is interesting to note that two of these systems
have a fast spin period: 69\,ms for \afive and 2.5\.ms for \sxj.

Assuming that the low luminosity emission is caused by centrifugal
inhibition of accretion, i.e. the so called ``propeller'' mechanism
at work in \afive, one would at first expect the spectrum to be
dominated by a thermal component from a residual accretion disk
up to the magnetospheric radius. Its absence is an indication
that indeed a pulsar mechanism may be at work driven by the
fast spinning neutron star.

\renewcommand{\refname}{\textbf{References}}
\parskip0pt
\bibsep0pt
\bibliographystyle{aa}
\bibliography{mnenomic,bexrb,accretion,diverse,crossref}

\begin{thebibliography}{20}
\expandafter\ifx\csname natexlab\endcsname\relax\def\natexlab#1{#1}\fi
\expandafter\ifx\csname url\endcsname\relax
  \def\url#1{{\tt #1}}\fi
\expandafter\ifx\csname urlprefix\endcsname\relax\def\urlprefix{URL }\fi

\bibitem[{{Alcock} et~al.(2001){Alcock}, {Allsman}, {Alves}
  et~al.}]{Alcock:2001}
{Alcock} C., {Allsman} R.A., {Alves} D.R., et~al.2001, MNRAS, 321, 678

\bibitem[{{Asai} et~al.(1998){Asai}, {Dotani}, {Hoshi} et~al.}]{Asai:98}
{Asai} K., {Dotani} T., {Hoshi} R., et~al.1998, PASJ, 50, 611

\bibitem[{{Brown} et~al.(1998){Brown}, {Bildsten}, \&
  {Rutledge}}]{BrownBildstenRutledge:98}
{Brown} E.F., {Bildsten} L., {Rutledge} R.E.1998, ApJL, 504, L95

\bibitem[{{Burderi} et~al.(2002){Burderi}, {Di Salvo}, {Stella}
  et~al.}]{Burderi:2002}
{Burderi} L., {Di Salvo} T., {Stella} L., et~al.2002, ApJ, 574, 930

\bibitem[{Campana(1997)}]{Campana:97}
Campana S.1997, A\&A, 320, 840

\bibitem[{{Campana} \& {Stella}(2000)}]{CampanaStella:2000}
{Campana} S., {Stella} L.2000, ApJ, 541, 849

\bibitem[{{Campana} et~al.(2002){Campana}, {Stella}, {Gastaldello}
  et~al.}]{Campana:2002}
{Campana} S., {Stella} L., {Gastaldello} F., et~al.2002, ApJL, 575, L15

\bibitem[{{Charles} et~al.(1983){Charles}, {Booth}, {Densham}
  et~al.}]{Charles:83}
{Charles} P.A., {Booth} L., {Densham} R.H., et~al.1983, MNRAS, 202, 657

\bibitem[{{Corbet} et~al.(1997){Corbet}, {Charles}, {Southwell}, \&
  {Smale}}]{Corbet:97}
{Corbet} R.H.D., {Charles} P.A., {Southwell} K.A., {Smale} A.P.1997, ApJ, 476,
  833

\bibitem[{Illarionov \& Sunyaev(1975)}]{IllarionovSunyaev:75}
Illarionov A.F., Sunyaev R.A.1975, A\&A, 39, 185

\bibitem[{{Johnston} et~al.(1980){Johnston}, {Griffiths}, \&
  {Ward}}]{Johnston:80}
{Johnston} M.D., {Griffiths} R.E., {Ward} M.J.1980, Nat, 285, 26

\bibitem[{{Mavromatakis} \& {Haberl}(1993)}]{MavromatakisHaberl:93}
{Mavromatakis} F., {Haberl} F.1993, A\&A, 274, 304

\bibitem[{{Pakull} \& {Parmar}(1981)}]{PakullParmar:81}
{Pakull} M., {Parmar} A.1981, A\&A, 102, L1

\bibitem[{{Ponman} et~al.(1984){Ponman}, {Skinner}, \& {Bedford}}]{Ponman:84}
{Ponman} T.J., {Skinner} G.K., {Bedford} D.K.1984, MNRAS, 207, 621

\bibitem[{{Skinner}(1980)}]{Skinner:80_optical}
{Skinner} G.K.1980, Nature, 288, 141

\bibitem[{{Skinner} et~al.(1980){Skinner}, {Shulman}, {Share}
  et~al.}]{Skinner:80_Xray}
{Skinner} G.K., {Shulman} S., {Share} G., et~al.1980, ApJ, 240, 619

\bibitem[{{Skinner} et~al.(1982){Skinner}, {Bedford}, {Elsner}
  et~al.}]{Skinner:82}
{Skinner} G.K., {Bedford} D.K., {Elsner} R.F., et~al.1982, Nature, 297, 568

\bibitem[{Stella et~al.(1994)Stella, Campana, Colpi, Mereghetti, \&
  Tavani}]{Stella:94}
Stella L., Campana S., Colpi M., Mereghetti S., Tavani M.1994, ApJ, 423, L47

\bibitem[{{White} \& {Carpenter}(1978)}]{WhiteCarpenter:78}
{White} N.E., {Carpenter} G.F.1978, MNRAS, 183, 11P

\bibitem[{Wijnands et~al.(2004)Wijnands, Heinke, Pooley
  et~al.}]{Wijnands:astro-ph}
Wijnands R., Heinke C., Pooley D., et~al.2004, submitted to ApJ.
  \texttt{astro-ph/0310144}

\end{thebibliography}

\end{document}